**Highlights**

- Sympathetic nerve activity is modulated by an interaction between central nervous system respiratory and sympathetic networks that leads to the generation of bursts of sympathetic nerve activity in phase with the respiratory cycle, termed respiratory-sympathetic coupling.

- In disease states altered respiratory sympathetic coupling may drive the development and/or maintenance of sympatho-excitation, and in turn hypertension, as is seen in chronic kidney disease.

- Using a rat model of chronic kidney disease, we showed in both young and adult animals the presence of enhanced respiratory-sympathetic coupling and an exaggerated sympathoexcitatory response to stimulation of the peripheral chemoreceptors with sodium cyanide (NaCN) and hypoxia, respectively.

- These results indicate that the pathways responsible for respiratory-sympathetic coupling, may be potential therapeutic targets to reduce high blood pressure in association with chronic kidney disease.

# Respiratory sympathetic modulation is augmented in chronic kidney disease

Manash Saha [a b c d], Clement Menuet [e f], Qi-Jian Sun [a], Peter G.R. Burke [g], Cara M. Hildreth [a], Andrew M. Allen [e], *Jacqueline K. Phillips [a]

[a] Department of Biomedical Sciences, Macquarie University, Australia

[b] Department of Nephrology, National Institute of Kidney Disease and Urology, Bangladesh

[c] Graduate School of Medicine, Wollongong University, Australia

[d] Department of Medicine, Wollongong Hospital, Australia

[e] Department of Physiology, University of Melbourne, Australia

[f] Institut de Neurobiologie de la Méditerranée, INMED UMR1249, INSERM, Aix-Marseille Université, Marseille, France

[g] Neuroscience Research Australia, Sydney NSW, Australia

*Running title*

Respiratory sympathetic coupling in CKD

*Author for correspondence

Professor Jacqueline Phillips

Department of Biomedical Science, Faculty of Medicine and Health Sciences

Macquarie University, Sydney NSW 2109 Australia

EMAIL: Jacqueline.phillips@mq.edu.au

PHONE: +61 2 9850 2753

**Abstract**

Respiratory modulation of sympathetic nerve activity (respSNA) was studied in a hypertensive rodent model of chronic kidney disease (CKD) using Lewis Polycystic Kidney (LPK) rats and Lewis controls. In adult animals under *in vivo* anaesthetised conditions *(n=8-10/strain)*, respiratory modulation of splanchnic and renal nerve activity was compared under control conditions, and during peripheral (hypoxia), and central, chemoreceptor (hypercapnia) challenge. RespSNA was increased in the LPK vs. Lewis (area under curve (AUC) splanchnic and renal: 8.7±1.1 vs. 3.5±0.5 and 10.6±1.1 vs. 7.1±0.2 µV.s, respectively, *P*<0.05). Hypoxia and hypercapnia increased respSNA in both strains but the magnitude of the response was greater in LPK, particularly in response to hypoxia. In juvenile animals studied using a working heart brainstem preparation *(n=7-10/strain)*, increased respSNA was evident in the LPK (thoracic SNA, AUC: 0.86±0.1 vs. 0.42±0.1 µV.s, *P*<0.05), and activation of peripheral chemoreceptors (NaCN) again drove a larger increase in respSNA in the LPK with no difference in the response to hypercapnia. Amplified respSNA occurs in CKD and may contribute to the development of hypertension.

1. **Introduction**

Hypertension is a major comorbidity associated with chronic kidney disease (CKD), arising early in the development of CKD and acting as a significant causal factor for the development of end-organ damage (Vanholder et al., 2005). Increased sympathetic nerve activity (SNA) is believed to play an important role in the development and/or maintenance of hypertension associated with kidney disease, with direct sympathetic nerve recording and plasma catecholamines levels elevated in individuals with CKD (Campese et al., 2011; Grassi et al., 2012; Klein IH, 2003; Phillips, 2005; Schlaich et al., 2009). Treatment with centrally acting sympatholytic agents can ameliorate hypertension in renal failure patients (Campese and Massry, 1983; Levitan et al., 1984; Schlaich et al., 2009), however, we still do not understand the mechanisms driving this increase in SNA.

Modulation of sympathetic nerve discharge occurs during the phases of respiration both in animals and humans, although the temporal relationship can differ between species and different nerve beds (Boczek-Funcke et al., 1992; Seals et al., 1993; Simms et al., 2009; Zoccal et al., 2008). This respiratory modulation of SNA (respSNA) creates synchronous alterations in blood pressure that allow optimal tissue perfusion. In experimental animals, where open-chest experiments have been performed, a component of the respSNA involves coupling between the relevant neuronal circuits within the brainstem (Machado et al., 2017; Simms et al., 2010). In the spontaneously hypertensive rat (SHR), a model of essential hypertension, respSNA is augmented and when compared to normotensive Wistar Kyoto control rats, the phase relationship is shifted from the post-inspiratory to inspiratory phase (Czyzyk-Krzeska and Trzebski, 1990; Simms et al., 2009). This exaggerated respSNA contributes to the development of hypertension in the SHR (Menuet et al., 2017). In CKD it is not known whether the phasic pattern of respSNA is altered or whether it contributes to the observed increase in SNA and the associated hypertensive state (Augustyniak et al., 2002; Salman et al., 2015).

The respSNA originating from central connections of respiratory and sympathetic networks (Haselton and Guyenet, 1989; Koshiya and Guyenet, 1996; Spyer, 1993; Sun et al., 1997) contributes independently to increased SNA in animal models following chronic intermittent hypoxia (Zoccal et al., 2008), with evidence of input from excitable central chemoreceptors (Molkov et al., 2011). The peripheral chemoreceptors may also provide a critical respiratory-related input that contributes to respSNA and, in disease states, drives the development and/or maintenance of sympatho-excitation, and in turn hypertension. This hypothesis is supported by a large body of work showing that exposure to chronic intermittent hypoxia (CIH) is correlated with enhanced respSNA (Machado et al., 2017). Patients with CKD are more vulnerable to respiratory disorders such as obstructive sleep apnoea (Hanly, 2004), and those with sleep apnoea *have elevated blood pressure* compared to those with CKD alone (Sekizuka et al.). Further, deactivation of the peripheral chemoreceptors, with hyperoxia, reduces muscle SNA in individuals with renal failure (Hering et al., 2007). In this context, there is exciting therapeutic potential in the investigation of the role of central respSNA and its underlying mechanism in hypertension CKD.

The main purpose of this study, therefore, was to test the hypothesis that respSNA is amplified in an animal model of CKD. Studies were undertaken in the Lewis polycystic kidney (LPK) rat, a genetic model of CKD presenting with kidney disease (McCooke et al., 2012) in which we have previously demonstrated hypertension, enhanced tonic SNA and perturbed reflex responses to both peripheral and central chemoreceptor stimulation (Salman et al., 2014; Salman et al., 2015; Yao et al., 2015).

2. Methods

All experimental procedures were approved by the Animal Ethics Committees of Macquarie University, NSW or the University of Melbourne, Victoria, Australia, and were carried out in accordance with the Australian Code of Practice for the Care and Use of Animals for Scientific Purposes.

*2.1 Study 1: Adult in vivo anaesthetised experiments*

Adult (12-13-week-old) male LPK ($n$ = 8) and control Lewis ($n$ = 10) rats were used. Animals were purchased from the Animal Resources Centre, Murdoch, Western Australia.

*2.1.1 Renal function*

A 24 h urine sample was collected from all animals 48 h prior to experimentation and urine volume, urinary creatinine and protein levels examined using a IDEXX Vetlab analyser (IDEXX Laboratories Pty Ltd., Rydalmere, NSW, Australia). At the commencement of the surgical procedure, an arterial blood sample was collected for determination of plasma urea and creatinine, and creatinine clearance calculated as described previously (Yao et al., 2015).

*2.1.2 Surgical procedures:*

Animals were anaesthetised with 10% (w/v) ethyl carbamate (Urethane, Sigma Aldrich, NSW, Australia) in 0.9% NaCl solution (1.3 g/kg i.p.). Reflex responses to hind-paw pinch were assessed to determine adequate depth of anaesthesia and supplemental doses of ethyl carbamate given as required (65 mg/kg i.p. or i.v.). Body temperature was measured and maintained at 37 ± 0.5° C using a thermostatically controlled heating blanket (Harvard Apparatus, Holliston, MA, USA) and infrared heating lamp. The right femoral vein and artery were cannulated for administration of fluid (Ringer's lactate, 5 ml/kg/h) and measurement of arterial pressure (AP) and blood collection for measurement of blood gases, respectively. The AP signal was sampled at 200 Hz and acquired using a CED 1401 plus and Spike2 software v.7 (Cambridge Electronic Designs (CED) Ltd, Cambridge, UK, RRID:SCR_000903). A tracheostomy was performed and an endotracheal tube placed *in situ*. A bilateral vagotomy was performed to cut afferent inputs from the stretch receptors of the lung and the animal was ventilated with oxygen enriched room air (7025 Rodent Ventilator, UgoBasile, Italy) and paralysed with pancuronium bromide (2 mg/kg iv for induction, 1 mg/kg for maintenance as required; AstraZeneca, North Ryde, NSW, Australia). The left phrenic, splanchnic and renal nerves were dissected and the distal end of each nerve was tied and cut. All nerves were bathed in a liquid paraffin pool and activity was recorded from the central end using bipolar silver wire recording electrodes. The activity was 10 times amplified, band-pass filtered between 10-1000Hz by a bio amplifier (CWE Inc., Ardmore, PA, USA) and sampled at 5kHz using CED 1401 plus and Spike2 software. All recordings were made with the same bioamplifier calibrated to a pre-set setting 50 µV.

Following surgical preparation, animals were then stabilised for 30 min. Arterial blood samples were collected and analysed for pH, $pCO_2$, $HCO_3^-$ and $SaO_2$ using a VetStat Electrolyte and Blood Gas Analyser (IDEXX Laboratories Pty Ltd., Rydalmere, NSW,

Australia). After the initial blood gas measurement, arterial blood gases were then corrected by adjusting the ventilator pump rate and volume and/or slow bolus injections of 5% sodium bicarbonate as required to maintain parameters under control conditions within the following range: pH = 7.4 ± 0.5; $PCO_2$ = 40 ± 5 mmHg; $HCO_3^-$ = 24 ± 2 mmol/L, $SaO_2$ = 100%; and end tidal (ET)$CO_2$ = 4.5 ± 0.5%. Repeat blood gas analysis was undertaken as required.

*2.1.3 Experimental protocol*

After the period of stabilization, the integrity of renal and splanchnic nerve recordings (rSNA and sSNA respectively) was confirmed by pulse modulation of SNA and demonstration of a baroreflex response to a bolus injection of phenylephrine (50 µg/kg iv, Sigma-Aldrich, St. Louis, MO, USA). Baseline parameters were measured under control conditions for a period of 30 min. Pilot studies indicated that hypoxic stimulation to stimulate the peripheral chemoreceptors using 10% $O_2$ in $N_2$ (Abbott and Pilowsky, 2009) could not be used for the 3 min stimulation period without affecting pH and $PCO_2$ in the LPK, and therefore animals were ventilated with room air, without oxygen supplementation, for a period of 3 min without any change in ventilator rate or volume. This induced a stable modest hypoxia - blood gas analysis parameters in the range of pH = 7.4 ± 0.5; $PCO_2$ = 43 ± 3 mmHg; $HCO_3^-$ = 24 ± 2 mmol/L and $SaO_2$ = 82-84%. After a 30 min recovery period, central chemoreceptor stimulation was performed by ventilating the animals with 5% carbogen (5% $CO_2$ in 95% $O_2$) for 3 min without any change in the ventilator rate or volume, which was confirmed by blood gas analysis to produce hypercapnia with parameters in the range of pH = 7.4 ± 0.5; $PCO_2$ = 62 ± 7 mmHg; $HCO_3^-$ = 24 ± 2 mmol/L and $SaO_2$ = 100%.

At the end of experiment, all animals were euthanized with potassium chloride (3M i.v.) and the electrical noise levels for phrenic nerve activity (PNA), sSNA and rSNA recorded and later subtracted for data analysis.

*2.1.4 Data analysis*

All data were analysed offline using Spike 2 software. From the AP signal, mean (MAP), systolic blood pressure (SBP), diastolic (DBP) and pulse (PP) pressure and heart rate (HR) were derived. The sSNA and rSNA were rectified and smoothed with a time constant of 0.1 s and PNA was rectified and smoothed with a time constant of 0.05 s. Baseline data was determined from a 30 s period at the end of the control ventilation period. Response to respiratory challenge was analysed over a 30 s period once PNA had stabilised for a period of 1-2 min and expressed as a change (Δ) relative to the 30 s period immediately prior to each stimulus. The PNA amplitude, frequency (number cycle.min$^{-1}$), duration (s) and minute activity (MPA=PNA amplitude x PNA frequency) were measured to quantify the changes. The phrenic cycle was divided into three phases: inspiratory (I), post-inspiratory (PI) and expiratory (E). For SNA, the mean level was measured from the period 200 ms prior to the onset of the phrenic burst, and this was considered the baseline for all other measurements. A phrenic-triggered average of the rectified and smoothed sSNA and rSNA recordings was performed and the phases divided into I, PI, and E. From this trace the following parameters were calculated for sSNA and rSNA: the peak amplitude (PA [μV]), the maximum amplitude at the apex of the peak of the SNA burst coincident with inspiratory/post-inspiratory phase, the duration (from onset of excitatory activity to return to baseline [s]) and area under curve (AUC) of respSNA excitatory peak (μV.s) determined as the integral of the waveform. AUC of respSNA for I, PI and E phase of phrenic cycle were also calculated from the integral of the waveform.

*2.2. Study 2: Juvenile working heart brainstem preparation*

To determine respSNA in early stages of the disease process, another series of experiments were performed in juvenile rats using the working heart brainstem in-situ preparation, which while a reduced preparation has the advantage of allowing assessment of physiological patterns of central respiratory control and cardiovascular regulation in the absence of the effects of anaesthesia (Wilson et al., 2001). Recordings were made in 5 week old male Lewis *(n = 7)* and LPK *(n = 10)* as described previously (Menuet et al., 2014). At this age, renal function is only mildly impaired in the LPK (Phillips et al., 2007). Animals were purchased from the Animal Resources Centre, Murdoch, Western Australia.

*2.2.1 Surgical procedure*

Animals were deeply anaesthetized with isoflurane until loss of the pedal withdrawal reflex and then dissected below the diaphragm, exsanguinated, cooled in Ringer solution on ice (composition in mM: 125 NaCl, 24 NaHCO3, 5 KCl, 2.5 CaCl2, 1.25 MgSO4, 1.25 KH2PO4 and 10 dextrose, pH 7.3 after saturation with carbogen gas (5% CO2, 95% O2) and decerebrated precollicularly. Lungs were removed and the descending aorta isolated and cleaned. Retrograde perfusion of the thorax and head was achieved via a double-lumen catheter (ø 1.25 mm, DLR-4, Braintree Scientific, Braintree, MA, USA) inserted into the descending aorta. The perfusate was Ringer solution containing Ficoll (1.25%) warmed to 31°C and gassed with carbogen (95% O2 and 5% CO2; closed loop reperfusion circuit). The second lumen of the cannula was connected to a transducer to monitor perfusion pressure in the aorta. Neuro-muscular blockade was achieved using vecuronium bromide added to the perfusate (2–4 μg/mL, Organon Teknika, Cambridge, UK). Simultaneous recordings of PNA, thoracic sympathetic nerve activity (tSNA, T8-10), vagus nerve activity (VNA) and abdominal nerve activity (AbNA, T9-T12) were obtained using glass suction electrodes. The activity was amplified (10 kHz, Neurolog), filtered (50–1500 kHz, Neurolog), digitized (CED) and recorded using Spike2 (CED).

*2.2.2 Experimental protocol*

After the period of stabilization, the peripheral chemoreceptors were stimulated using sodium cyanide (NaCN; 0.05%; 100 μL bolus) injected into the aorta via the perfusion catheter (Chang et al., 2015; Menuet et al., 2016). Central chemoreceptors were stimulated using hypercapnic conditions by changing the level of $CO_2$ in the perfusate source from 95% O2 and 5% $CO_2$ to 90% O2 and 10% $CO_2$ for ~5 min. The electrical noise levels for tSNA recordings were determined at the end of experiments by sectioning the sympathetic chain at the proximal paravertebral ganglion level and were subtracted for data analysis. All chemicals were purchased from Sigma-Aldrich, Australia.

*2.2.3 Data analysis*

All data were analysed offline using Spike 2 software. The signals were rectified and integrated with a 50 ms time constant. The HR was derived using a window discriminator to trigger from the R-wave of the electrocardiogram recorded simultaneously with PNA. The PNA, and VNA signals were used to assess respiratory parameters associated with inspiration, post-inspiration and expiration, respectively.

During baseline and hypercapnic conditions, phrenic-triggered (end of inspiratory burst) averaging of tSNA was used for the analysis of tSNA parameters related to the burst of respSNA tSNA and tonic non-respiratory modulated tSNA (baseline tSNA). NaCN-induced respSNA was measured as delta increase in AUC compared to a pre-stimulus control period with same duration (Menuet et al., 2016), while peak activity of respSNA of tSNA was calculated as the maximum amplitude of the tSNA burst coincident with inspiratory/post-inspiratory transition.

*2.3 Statistical analysis*

All data are presented as mean ± SEM. A Student's t-test was used to determine baseline differences in cardiorespiratory function and respiratory sympathetic coupling between the LPK and Lewis. A two-way ANOVA with repeated measures followed by Bonferroni's post-hoc analysis was used to analyse the effect of peripheral or central chemoreceptor challenge on respiratory sympathetic coupling with strain and chemoreceptor challenge as the variables. All analysis was performed using GraphPad Prism software v6.0 (GraphPad Software Inc., La Jolla, California, USA, RRID:SCR_002798). Differences were considered statistically significant where $P<0.05$.

3. Results

*3.1 Adult baseline parameters*

Adult male LPK rats showed a phenotypic elevation in blood urea (24.3 ± 2.3 vs. 5.9 ± 0.3 mmol/L) and plasma creatinine (60.6 ± 11.7 vs. 22.3 ± 4.6 μmol/L) and reduction in creatinine clearance (2.08 ± 0.4 vs. 10.1 ± 2.3 mL/min) (all LPK *n*=8 vs. Lewis *n*=10; $P< 0.001$) reflective of impaired renal function.

During the control period, individual recordings were made of rSNA, sSNA, PNA and AP from Lewis and LPK rats (Figure 1). Consistent with our previous work (Salman et al., 2014; Yao et al., 2015), SBP, MAP, DBP, PP, HR and SNA (both sSNA and rSNA) were elevated in the adult LPK compared with age-matched Lewis controls (Table 1). Phrenic frequency was also significantly higher in the adult LPK, suggestive of increased central respiratory drive; however, there was no significant difference in PNA amplitude or MPA between strains (Table 1).

We examined the temporal relationship between SNA and the respiratory cycle in LPK and Lewis rats under control conditions. Both the rSNA and sSNA exhibited distinct respSNA, with a clear burst in SNA in the PI period in both the Lewis and LPK (see Figures 2 and 4, control panels A -B [renal], E-F [splanchnic]). In the Lewis rat, both sSNA and rSNA showed a tendency to increase from baseline during I, before the sharp peak occurred in PI. After the PI peak, activity fell quickly to baseline and was relatively flat during E. The LPK showed a distinct difference – with a decrease in activity below baseline during early I that was most obvious in rSNA (rSNA, I phase AUC, LPK vs Lewis: -0.07 ± 0.03 vs 0.45 ± 0.2; $P \leq 0.5$; sSNA, I phase AUC, LPK vs Lewis: 0.06 ± 0.05 vs 0.18 ± 0.08 ; $P > 0.5$). There was a gradual decline in both sSNA and rSNA during the E period which was not significantly different between the strains (rSNA, E phase AUC, LPK vs Lewis: 0.07 ± 0.04 vs 0.08 ± 0.1; $P > 0.5$; sSNA, E phase AUC, LPK vs Lewis: 0.02 ± 0.03 vs 0.04 ± 0.02; $P > 0.5$ [Figures 2 and 4, control panels A -B (renal), E-F (splanchnic)].

Quantitative parameters characterizing respSNA under control conditions for both nerves are provided in Table 2. PA and AUC for both sSNA and rSNA were significantly greater in the LPK compared with Lewis. The duration of the respSNA for both sSNA and rSNA was not different between the Lewis and LPK indicating that the greater AUC observed in the LPK was driven primarily by the greater PA.

*3.2 Adult responses to chemoreceptor stimulation*

In response to a hypoxic challenge, a greater increase in AP was observed in the LPK compared with Lewis (Table 3). In the LPK rat this was associated with a slight, but significant, slowing of PNA frequency, but not amplitude or duration. The hypercapnic challenge produced comparable increases in AP and central respiratory drive in the two strains, with only a small difference in the HR response.

In both strains, under hypoxic and hypercapnic conditions, the respSNA curves of sSNA and rSNA retained a consistent PI peak (see Figures 2 and 4). In response to hypoxia, respSNA (sSNA and rSNA) was increased in both strains as reflected by an increase in PA in Lewis and an increase in PA and AUC in LPK (Figure 3) that was greater in LPK rats (delta AUC, sSNA: Lewis vs. LPK: 0.8 ± 0.6 vs. 5.9 ± 1.9, rSNA: 2.8 ± 0.7 vs. 7.5 ± 2.1 µV.s, both *P< 0.05*). Hypercapnia induced an increase in the respSNA of sSNA as reflected by an increase in PA in Lewis and an increase in PA and AUC in LPK. In the rSNA there was an increase in PA in the Lewis only (Figures 4, 5). However, the magnitude of the change in the respSNA of both nerves was similar between strains (Delta AUC, sSNA, Lewis vs. LPK: 1 ± 0.4 vs. 2.5 ± 0.7; rSNA, Lewis vs. LPK: 3.2 ± 0.5 vs. 2.5 ± 1.4 µV.s, both *P > 0.05*). During hypoxia, the inspiratory inhibition in the LPK increased in rSNA (I phase AUC, rSNA, control vs. hypoxia: -0.07 ± 0.04 vs. -0.9 ± 0.3 µV.s p < 0.05; Figure 2) and became clearly evident in sSNA (I phase AUC, sSNA, control vs. hypoxia : 0.05 ± 0.04 vs. -0.4 ± 0.1 µV.s *P < 0.05*).

This increase in inspiratory inhibition did not occur in the LPK under hypercapnic conditions ( I phase AUC, rSNA, control vs. hypercapnia : -0.32 ± 0.2 vs. -0.4 ± 0.2 sSNA: 0.03 ± 0.09 vs. 0.06 ± 0.1 µV.s both *P >0.05*). There was no significant change to the inspiratory phase in the Lewis animals under either condition.

### *3.3 Juvenile baseline parameters*

In the working heart brainstem preparation (Figure 6), mean perfusion pressure was not different between the two strains (Lewis vs. LPK: 73 ± 4 vs. 76 ± 2 mmHg, *P* = 0.55). The HR was faster in the LPK (Lewis vs. LPK: 360 ± 13 vs. 302 ± 17 bpm, *P* < 0.05). There was no difference observed in PNA frequency (Lewis vs. LPK: 9.5 ± 0.7 vs. 10.2 ± 1.6, *P = 0.74*), although inspiratory duration of PNA was reduced in the LPK (Lewis vs. LPK: 1658 ± 147 vs. 1242 ± 79 ms, *P* < 0.05). Tonic levels of tSNA were not significantly different (Lewis vs. LPK: 4.37 ± 0.47 vs. 4.51 ± 0.58 µV, *P = 0.85*); however, respiratory modulation of tSNA was greater in the LPK with a larger AUC observed (Lewis vs. LPK: 0.42 ± 0.13 vs. 0.86 ± 0.13 µV.s, *P* < 0.05). No differences were observed in respSNA burst duration (Lewis vs. LPK: 0.8 ± 0.1 vs. 1.4 ± 0.2 s, *P = 0.10*) or the magnitude of the peak amplitude of tSNA (Lewis vs. LPK: 6.01 ± 0.56 vs. 6.17 ± 0.85 µV, *P = 0.87*).

### *3.4 Juvenile responses to chemoreceptor stimulation*

Peripheral chemoreceptor activation with NaCN produced a comparable increase in perfusion pressure in both strains (Lewis vs. LPK: 5.7 ± 1.4 vs. 6.4 ± 1.2 mmHg, *P = 0.71*). The accompanying large bradycardia was of greater magnitude in the LPK (Lewis vs. LPK: -86 ± 18 vs. -176 ± 16 bpm, *P* < 0.05). RespSNA increased in both strains in response to peripheral chemoreceptor activation (NaCN, *P<0.05*; Figure 6), as evidenced by an increase in PA and AUC that was of greater magnitude in both circumstances in the LPK. (Lewis vs. LPK: delta PA: 3.6 ± 0.6 vs. 6.5 ± 0.9 µV, delta AUC: Lewis vs. LPK: 7.0 ± 1.2 vs. 15.1 ± 2.7 µV.s, *P<0.05*).

Central chemoreceptor activation with hypercapnia produced a comparable reduction in perfusion pressure (Lewis vs. LPK: -7.9 ± 1.9 vs. -8.6 ± 1.1 mmHg, *P = 0.75*) and HR (Lewis vs. LPK: -42 ± 9 vs. -53 ± 8 bpm, *P = 0.36*) in the two strains. RespSNA increased in both strains (Figure 7) with an increase in PA in the LPK and a comparable increase in AUC in both strains (delta AUC: 0.98 ± 0.16 vs. 0.99 ± 0.29 µV.s, *P = 0.97*).

## 4. Discussion

Our study has determined the characteristic features of respSNA in a rodent model of CKD. We have directly measured the relationship between inspiratory drive and two sympathetic nerves in anaesthetized adult animals and to one sympathetic nerve in juvenile animals using the working heart brainstem preparation. The timing of the peak of respSNA was observed persistently in the PI period under all conditions tested and, from the earliest age studied (5 weeks). Adult and juvenile LPK rats demonstrated increased respSNA compared to Lewis control rats. Under control conditions, the LPK rat shows an alteration in respSNA patterning with increased inhibition of rSNA during the inspiratory period. This inspiratory inhibition is augmented in rSNA with hypoxia and becomes evident in sSNA with this respiratory challenge. In addition, we have found that peripheral chemoreceptor stimulation has an exaggerated effect on respSNA in both the juvenile and adult LPK animals, which suggests a role for peripheral chemoreceptors in the pathophysiology of autonomic dysfunction associated with CKD.

Our finding of augmented respSNA in the LPK model of CKD is consistent with previous work in the SHR model, which also demonstrates amplified respSNA when examined using the same methodologies in juvenile (Menuet et al., 2017; Simms et al., 2009) and adult SHRs (Czyzyk-Krzeska and Trzebski, 1990) that we have applied, though notably, we do not see the marked temporal shift in the peak response from post-inspiration to inspiration that is described in the SHR during the development of hypertension. Enhancement of respSNA is also a feature noted in animal models of chronic intermittent hypoxia (Zoccal et al., 2008), congestive heart failure (Marcus et al., 2014); and in rats following uteroplacental insufficiency (Menuet et al., 2016). Common to these diseases is autonomic dysfunction and the development of hypertension in association with increased SNA.

While we did not see elevated baseline tSNA or perfusion pressures in the juvenile LPK at 5 weeks of age, our previous work in both young and adult LPK (7 and 14 weeks) show increased SNA, including recordings from conscious animals (Salman et al., 2014; Salman et al., 2015), and we observe markedly elevated blood pressure in these animals from an early age (6 weeks) (Phillips et al., 2007; Sarma et al., 2012). Our demonstration of increased respSNA at age 5 weeks in the current study indicates that altered sympathetic activity is a key feature of the autonomic dysfunction in the LPK model, consistent with clinical data demonstrating that in patients with polycystic kidney disease, muscle SNA is elevated very early in the disease, often prior to any significant reduction in renal function (Klein et al., 2001). Our work is supportive of the hypothesis that in CKD, altered interactions between central respiratory and sympathetic pathways is a contributor to the development and maintenance of hypertension.

The rostral ventrolateral medulla (RVLM) is an important site in the generation and regulation of sympathetic vasomotor tone (Guyenet, 2006), and our recent work in the SHR suggests that the inspiratory pre-Bötzinger complex is a likely site of modulatory inputs to the RVLM C1 neurons which could drive an increase in SNA (Menuet et al., 2017). There is also an important contribution from the respiratory pattern generator that results in phasic modulation of RVLM presympathetic neurons. Recordings from both anesthetised rats and juvenile rats using the working heart brainstem preparation show 3 main classes of respiratory-modulated RVLM presympathetic neurons – I activated, PI activated and I inhibited – as well as some none modulated neurons (Haselton and Guyenet, 1989; Moraes et al., 2013). Whilst not directly tested in this study, we propose that altered strength of respiratory input to these different neuron classes could underlie the different patterns of respSNA observed including the enhanced I inhibition of respSNA in the LPK rat.

While the exact mechanisms underlying amplified respSNA have not yet been elucidated, it has been suggested that interaction between central (Molkov et al., 2011) and peripheral chemoreceptor pathways (Guyenet et al., 2009; Moraes et al., 2015) may contribute to increased sensitivity of retrotrapezoid nucleus (RTN). The RTN has been shown to play a critical role in amplified respSNA under conditions of chronic intermittent or sustained hypoxia (Molkov et al., 2014; Moraes et al., 2014). Of relevance to our work is that peripheral chemoreceptor hypersensitivity has been suggested to contribute to the development of sympathetic over activity in kidney disease (Hering et al., 2007). In the current study, stimulation of peripheral chemoreceptors by NaCN and hypoxia evoked an increase in respSNA parameters in both the juvenile and adult animals, respectively, that was of a significantly greater magnitude in the LPK. In contrast, exposure to hypercapnia, which drives a centrally mediated chemoreflex and subsequent sympathoexcitatory and pressor response, produced changes which were similar between the strains. Moreover, adult LPK rats showed a higher respSNA even in control condition, and inspiratory motor activity was similar between strains after both peripheral and central chemoreceptor stimulation. Given the evidence that chronic intermittent hypoxia induces sensitization of peripheral chemoreceptors, producing an exaggerated sympathoexcitation and increased expiratory-related SNA (Braga et al., 2006; Moraes et al., 2012; Zoccal et al., 2008), and that increased peripheral chemoreceptor sensitivity contributes to sympathetic over activity and hypertension in SHR rats (Paton et al., 2013b; Tan et al., 2010), we speculate that there may be at least two mechanisms working centrally in the medulla oblongata to contribute the augmented respSNA seen in association with hypertension in CKD; one dependent on peripheral chemoreceptors and another one that is independent.

Hypoxia is a common contributing factor for amplified respSNA and increased SNA in disease conditions including hypertension (Calbet, 2003; Somers et al., 1988; Zoccal et al., 2008), and while we did not investigate directly the mechanism behind increased respSNA in CKD in the present study, of relevance is our previous published data that has documented vascular remodelling, hypertrophic changes and calcification in the aortic arch of LPK animals (Salman *et al*., 2014). This type of vascular remodelling is closely associated with ageing, hypertension and CKD in humans and is documented to occur in the carotid/vertebral arteries, and in association with atherosclerosis, can result in increased intima-media thickness and ultimately narrowing the lumen of the carotid and vertebral artery (Tanaka et al., 2012; Yamada et al., 2014). Given that the carotid body and brainstem are highly vascular organs and peripheral/central chemoreceptors are sensitive to reduced blood flow, remodelling of the carotid body/cerebral arterioles could be a persistent stimulus for central/peripheral chemoreceptor hypersensitivity (Moraes et al., 2014; Paton et al., 2013a). Anaemia is also common in CKD, a feature well documented in the LPK strain (Phillips et al., 2015), which in itself may be cause of persistant hypoxia. Regardless of the mechanism, we presume that in CKD, a persistent hypoxic stimulus via peripheral chemoreceptor independent and/or dependent pathways, can drive premotor sympathetic neurons and respiratory (expiratory) neurons to produce amplified respSNA and increased sympathetic tone in similar way to that of other pathological conditions (Wong-Riley et al., 2013; Zoccal and Machado, 2010; Zoccal et al., 2008).

In conclusion, we provide evidence that respSNA is augmented and associated with increased SNA and hypertension in CKD. We further demonstrate an exaggerated respSNA response to peripheral chemoreceptor stimulation, indicating this modulatory pathway may act as one of the drivering factors of autonomic dysfunction in CKD. The pathways responsible for modulation of respSNA, from the carotid body chemoreceptors to central sites such as the RVLM, may well be valuable novel therapeutic targets in CKD. Our data also show that changes in respSNA are evident very early in the disease process, suggesting that early intervention may help to reduce the complex pathogenesis of CKD.

## Funding sources

This work was supported by the National Health and Medical Research Council of Australia (GNT1030301, GNT1030297, GNT1102477) and Macquarie University

Australia. M Saha is a recipient of a Macquarie University International Research Scholarship, and M Menuet was supported by a McKenzie Research Fellowship from the University of Melbourne, Australia.

# Tables

*Table 1: Baseline cardiorespiratory function in adult Lewis and LPK rats*

|  | Lewis | LPK |
|---|---|---|
| MAP (mmHg) | 90 ± 4 | 125 ± 10* |
| SBP (mmHg) | 117 ± 8 | 194 ± 22* |
| DBP (mmHg) | 75 ± 3 | 99 ± 10* |
| PP (mmHg) | 41 ± 6 | 87 ± 18* |
| HR (bpm) | 453 ± 9 | 480 ± 3* |
| PNA amplitude (µV) | 18.3 ± 2.1 | 24.9 ± 5.3 |
| PNA frequency (cycles/min) | 36 ± 1 | 45 ± 1* |
| PNA duration (s) | 0.82 ± 0.1 | 0.72 ± 0.01 |
| MPA | 674.9 ± 86.03 | 1125 ± 232 |
| sSNA (µV) | 3.1 ± 0.5 | 7.9 ± 1.3* |
| rSNA (µV) | 5.2 ± 0.4 | 8.9 ± 0.6* |

Measures of cardiorespiratory function in Lewis and LPK rats under control conditions. MAP: mean arterial pressure, SBP: systolic blood pressure, DBP: diastolic blood pressure, PP: pulse pressure, HR: heart rate, bpm: beats per min, PNA: phrenic nerve amplitude, MPA: minute phrenic activity, rSNA: renal sympathetic nerve activity, sSNA: splanchnic sympathetic nerve activity. Results are expressed as mean ± SEM. *P<0.05* between the Lewis and LPK as determined by Student's t-test. *n* = 9 Lewis and n = 8 LPK excepting rSNA where n = 5 LPK and 6 Lewis

*Table 2: RespSNA parameters in splanchnic and renal sympathetic nerves under control conditions.*

|  |  | Lewis | LPK |
|---|---|---|---|
| PA (μV) | splanchnic | 4.1 ± 0.8 | 8.8 ± 1.1* |
|  | renal | 8.7 ± 0.6 | 11.02 ± 0.8* |
| Duration (sec) | splanchnic | 0.8 ± 0.07 | 0.99 ± 0.02 |
|  | renal | 0.8 ± 0.05 | 0.96 ± 0.05 |
| AUC (μV.s) | splanchnic | 3.5 ± 0.5 | 8.7 ± 1.1* |
|  | renal | 7.1 ± 0.2 | 10.6 ± 1.1* |

Measures of respSNA (phrenic triggered) parameters for integrated sSNA and rSNA in Lewis and LPK rats under control conditions. PA: peak amplitude, AUC: area under curve. Results are expressed as mean ± SEM. *P<0.05* between the Lewis and LPK as determined by Student's t-test. sSNA data: *n* = 9 Lewis and n = 8 LPK, excepting rSNA where n = 5 LPK and 6 Lewis.

*Table 3: Effects of peripheral and central chemoreceptor stimulation on cardiorespiratory parameters in adult Lewis and Lewis Polycystic Kidney (LPK) rats*

|  |  | Lewis (n = 9) | LPK (n = 8) |
|---|---|---|---|
| Hypoxia | Δ MAP (mmHg) | 4 ± 4 | 21 ± 5 * |
|  | Δ SBP (mmHg) | 5 ± 5 | 34 ± 11 * |
|  | Δ DBP (mmHg) | 4 ± 4 | 16 ± 4 |
|  | Δ PP (mmHg) | 1 ± 2 | 18 ± 7 * |
|  | Δ HR (bpm) | 12 ± 2 | 10 ± 1 |
|  | Δ PNA amplitude (μV) | 10.9 ± 3.6 | 9.8 ± 3.9 |
|  | Δ PNA duration (sec) | 0.13 ± 0.02 | 0.07 ± 0.01 |
|  | Δ PNA frequency (cycles/min) | 5 ± 2 | -2 ± 2* |
|  | Δ MPA | 536.6 ± 152.5 | 300.7 ± 66.9 |
| Hypercapnea | Δ MAP (mmHg) | 19 ± 3 | 16 ± 4 |
|  | Δ SBP (mmHg) | 22 ± 4 | 29 ± 8 |
|  | Δ DBP (mmHg) | 18 ± 3 | 13 ± 3 |
|  | Δ PP (mmHg) | 4 ± 2 | 16 ± 6 |
|  | Δ HR (bpm) | -8 ± 1 | -4 ± 1 * |

| | | |
|---|---|---|
| Δ PNA amplitude (µV) | 8.7 ± 2.7 | 9.1 ± 1.6 |
| Δ PNA duration (sec) | 0.15 ± 0.02 | 0.11 ± 0.02 |
| Δ PNA frequency (cycles/min) | -3 ± 1 | -4 ± 1 |
| Δ MPA | 178.5 ± 74.08 | 293.9 ± 71.04 |

Delta change in phrenic nerve activity (PNA) and blood pressure (mmHg) under hypoxic or hypercapnic conditions in adult Lewis and Lewis Polycystic Kidney (LPK) rats. MAP: mean arterial pressure, SBP: systolic blood pressure, DBP: diastolic blood pressure, PP: pulse pressure; HR: heart rate, MPA: minute phrenic activity. Results are expressed as mean ± SEM. *P<0.05* between the Lewis and LPK as determined by Student's t-test. Δ = Delta change in, n = number of animals per group

# Figures

*Figure 1. Representative traces showing higher sympathetic nerve activity during different phases of respiration in adult Lewis and LPK rats*

Representative data traces showing raw and integrated splanchnic SNA (sSNA and ∫sSNA), raw and integrated renal SNA (rSNA and ∫rSNA), raw phrenic nerve activity (PNA) and raw arterial pressure (AP) in an adult Lewis (A) and LPK rat (B) under control conditions. Traces illustrate the higher rSNA, sSNA and AP including pulse pressure range in the LPK in comparison to Lewis rats. Expanded trace shows raw and integrated splanchnic and renal SNA over two respiratory cycles, noting different scale for floating scale bars for Lewis and LPK to allow for clear illustration of respSNA.

*Figure 2. Tracings of effect of hypoxia on respiratory-related sympathetic nerve activity in adult Lewis and LPK rats.*

Example figure illustrating phrenic triggered integrated rSNA (∫rSNA) and sSNA (∫sSNA) during different phases of the phrenic cycle (I, inspiration; PI, post inspiration; E, expiration within trace of phrenic nerve activity PNA) under control (panels A, B, E, F) and hypoxic conditions (panels C, D, G, H) in a Lewis rat (panels A, C, E, G) and LPK rat (panels B, D, F, H).

*Figure 3. Grouped data of effect of hypoxia on respiratory-related sympathetic nerve activity in adult Lewis and LPK rats.*

Grouped data is shown in panels A-D illustrating the change in peak amplitude (μV, A/B), and area under the curve ((μV.s, AUC; C/D) of the phrenic triggered ∫sSNA (panels A/C) and ∫rSNA (panels B/D) under control ($SaO_2$=100%) and hypoxic ($SaO_2$=82-84%) conditions. Data is expressed as mean ± SEM n ≥5 per group. * represents *P<0.05* versus control within each strain, # represents *P<0.05* versus treatment-matched Lewis.

*Figure 4. Tracings of effect of hypercapnia on respiratory-related sympathetic nerve activity in adult Lewis and LPK rats.*

Example figure illustrating phrenic triggered integrated rSNA (∫rSNA) and sSNA (∫sSNA) during different phases of the phrenic cycle (I, inspiration; PI, post inspiration; E, expiration within trace of phrenic nerve activity PNA) under control (panels A, B, E, F) and hypercapnic (panels C, D, G, H) conditions in a Lewis rat (panels A, C, E, G) and LPK rat (panels B, D, F, H).

*Figure 5. Group data of effect of hypercapnia on respiratory-related sympathetic nerve activity in adult Lewis and LPK rats.*

Grouped data is shown in panels A-F illustrating the change in peak amplitude (A/B), area under the curve (AUC; C/D) and duration (E/F) of the phrenic triggered ∫sSNA (panels A/C/E) and ∫rSNA (panels B/D/F) under control and hypercapnic conditions. Data is expressed as mean ± SEM n ≥5 per group. * represents *P<0.05* versus control within each strain # represents *P<0.05* versus treatment-matched Lewis.

*Figure 6. Effect of stimulation of the peripheral chemoreceptor reflex with NaCN on respiratory-related sympathetic nerve activity in juvenile Lewis and LPK rats using the working heart brainstem preparation.*

Representative figure illustrating integrated tSNA (∫tSNA) over the respiratory cycle in a Lewis rat (panel A) and LPK rat (panel B), noting different floating scale bars for Lewis and LPK (∫VNA) to allow for clear illustration of nerve activity. Grouped data illustrating effect of peripheral chemoreceptor stimulation on peak amplitude (C) and area under the curve (AUC; D) of phrenic triggered ∫tSNA. Data is expressed as mean ± SEM n minimum 5 per group. * represents *P<0.05* versus control within each strain # represents *P<0.05* versus treatment-matched Lewis.

*Figure 7. Effect of stimulation of the central chemoreceptor reflex with hypercapnia on respiratory-related sympathetic nerve activity in juvenile Lewis and LPK rats using the working heart brainstem preparation.*

Representative figure illustrating phrenic triggered integrated tSNA (∫tSNA) over the respiratory cycle in a Lewis rat (panels A and B) and LPK rat (panels C and D) under control (panels A and C) and hypercapnic (panels B and D) conditions. Grouped data illustrating the effect of central chemoreceptor stimulation on peak amplitude (E) and area under the curve (AUC; F) of the phrenic triggered ∫tSNA. Data is expressed as mean ± SEM n ≥5 per group. * represents *P<0.05* versus control within each strain, # represents *P<0.05* versus treatment-matched.

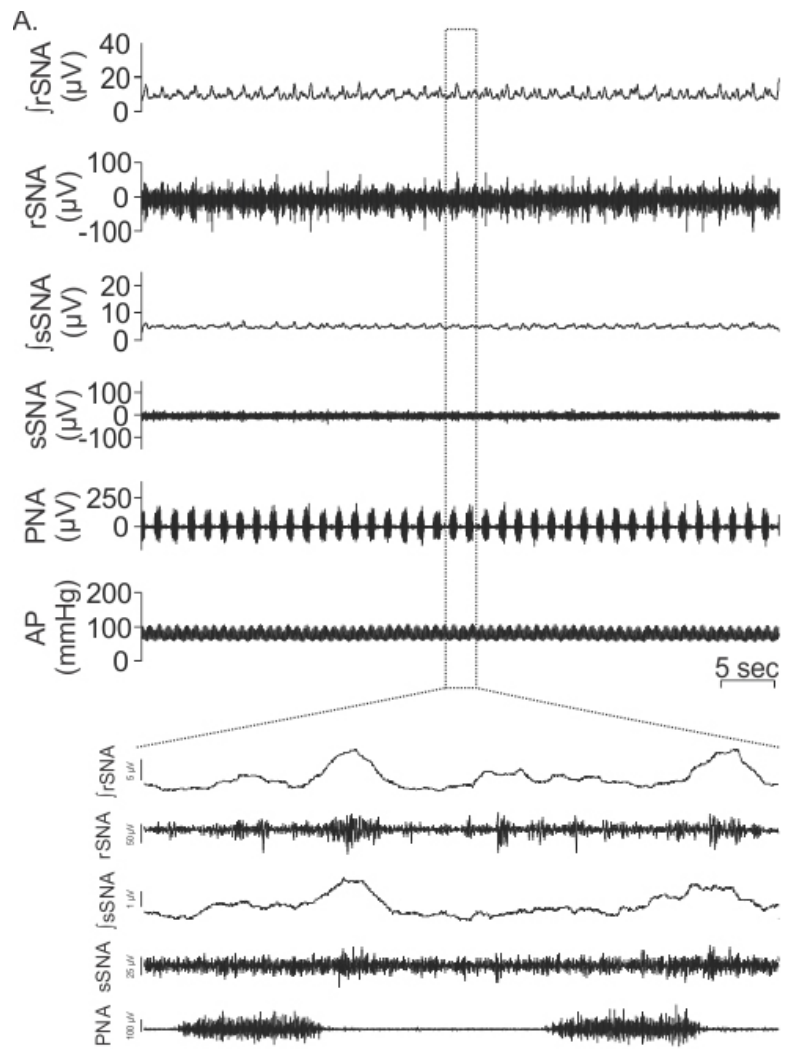
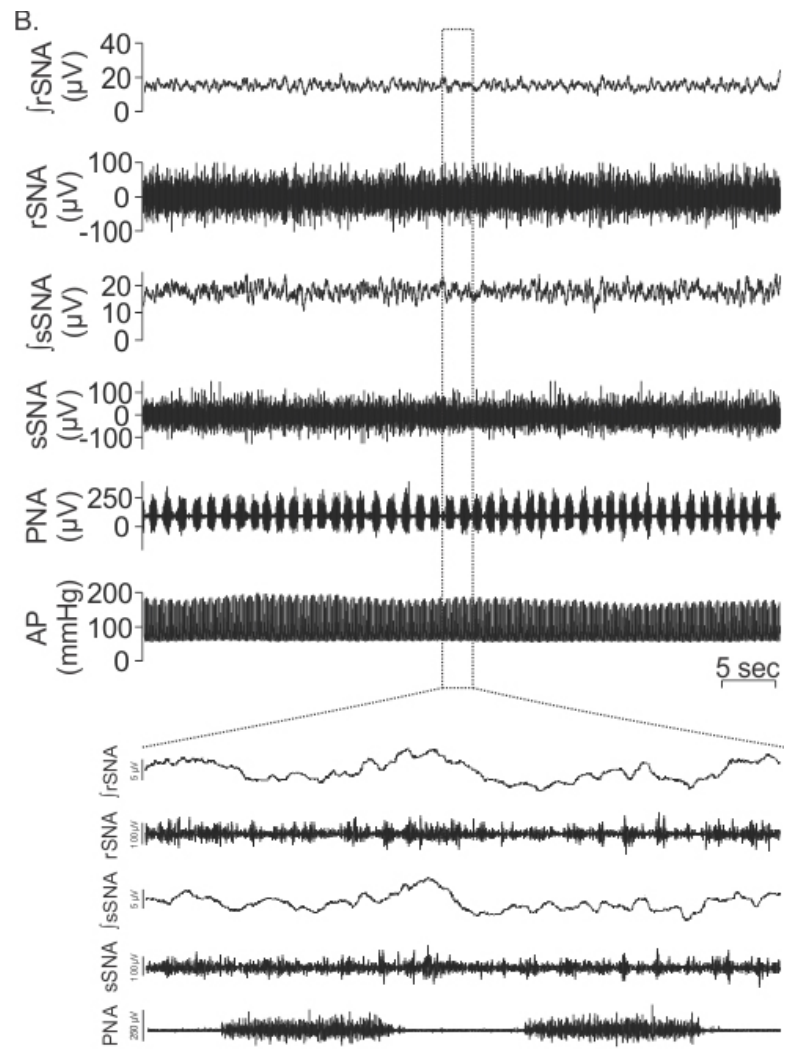

Figure 1

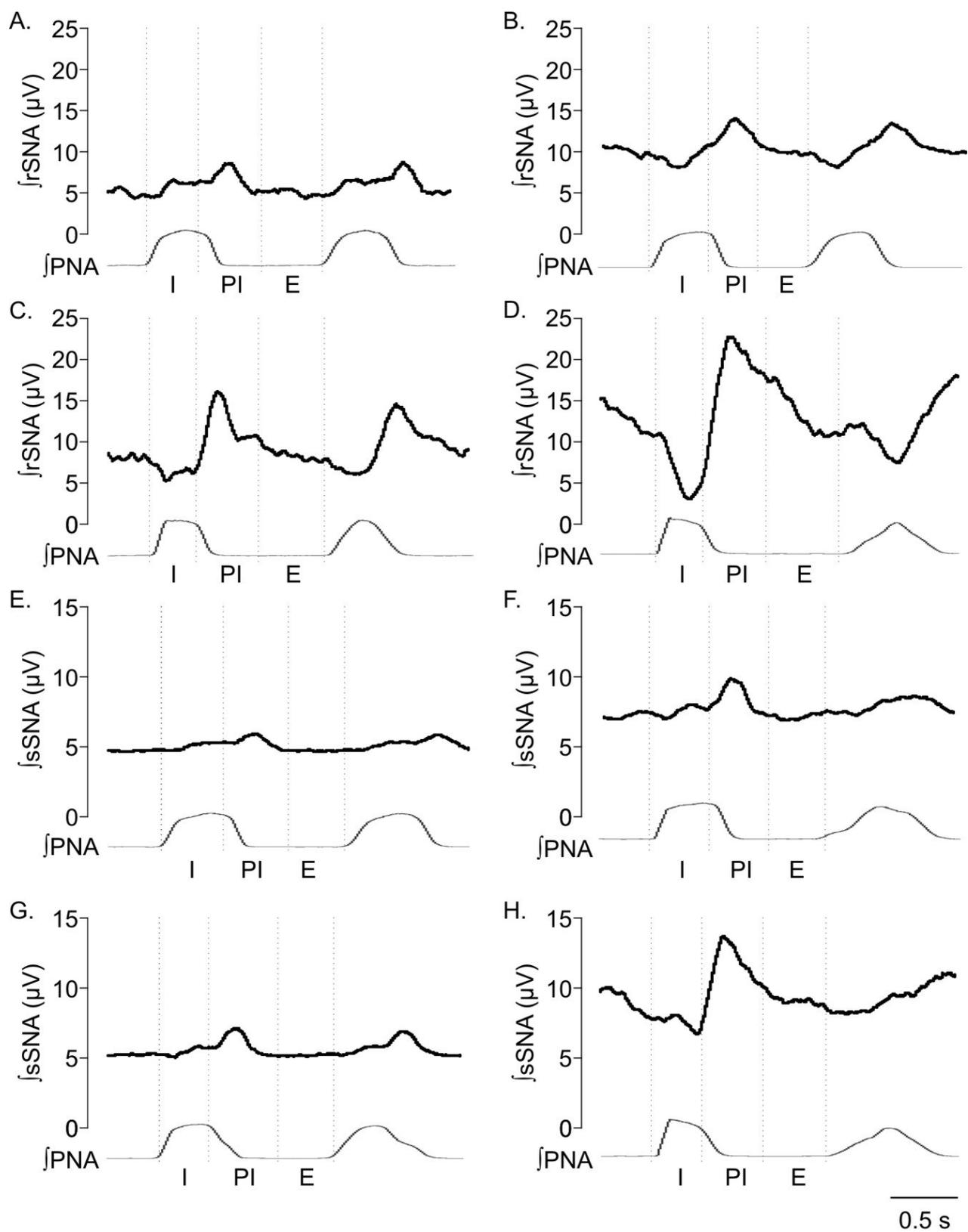

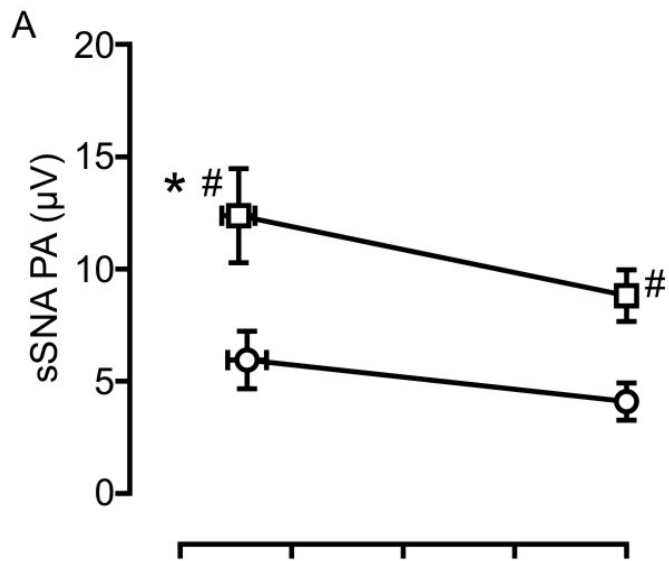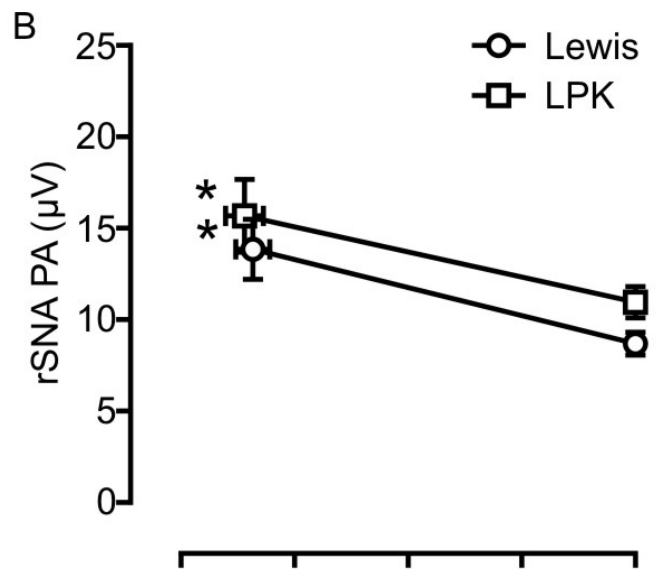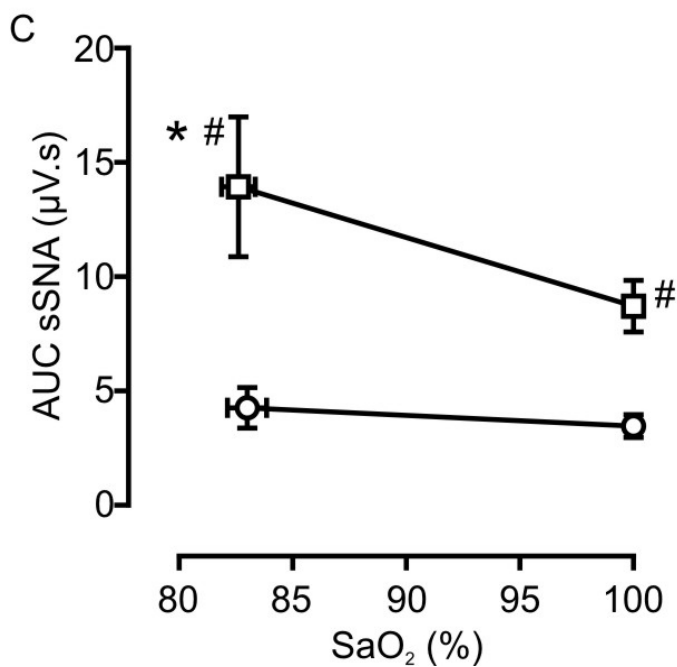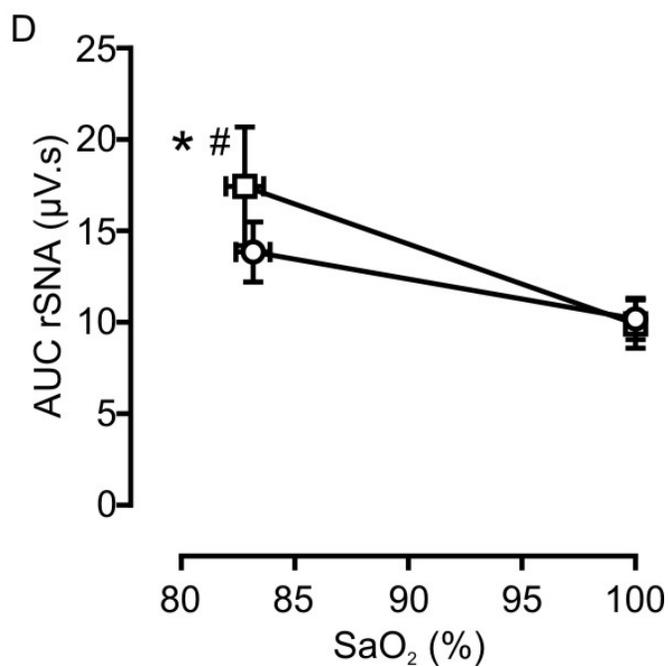

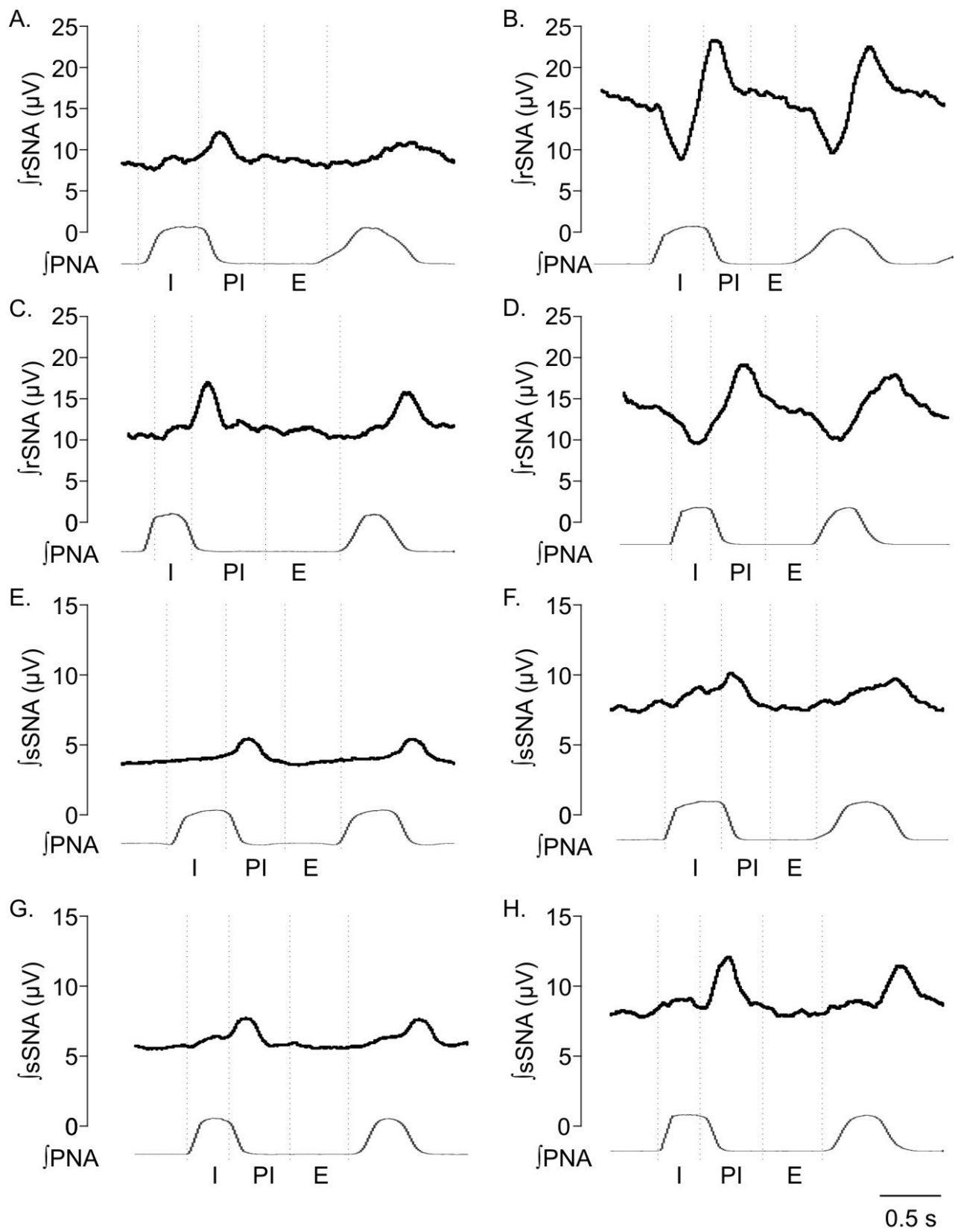

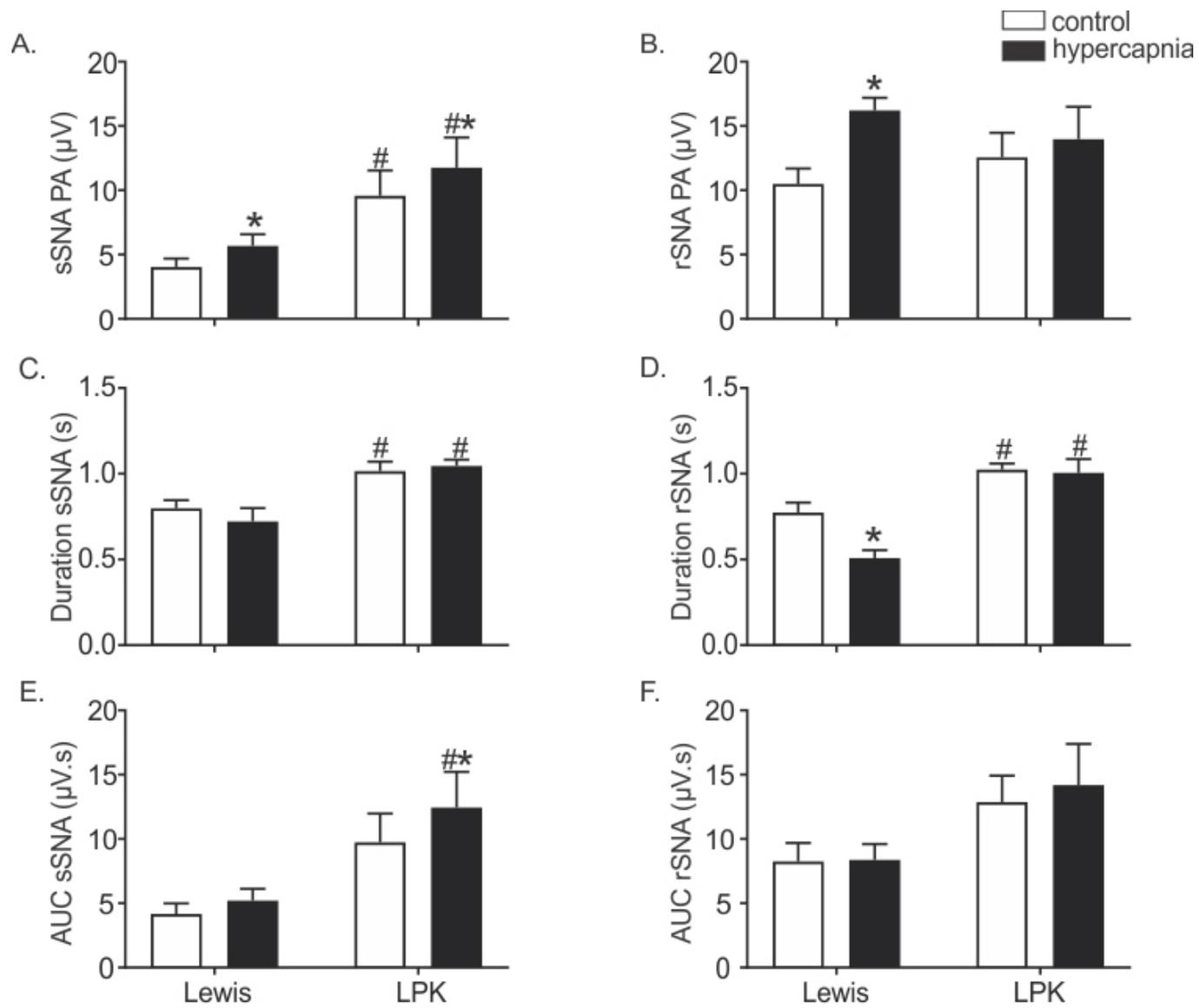

Figure 5

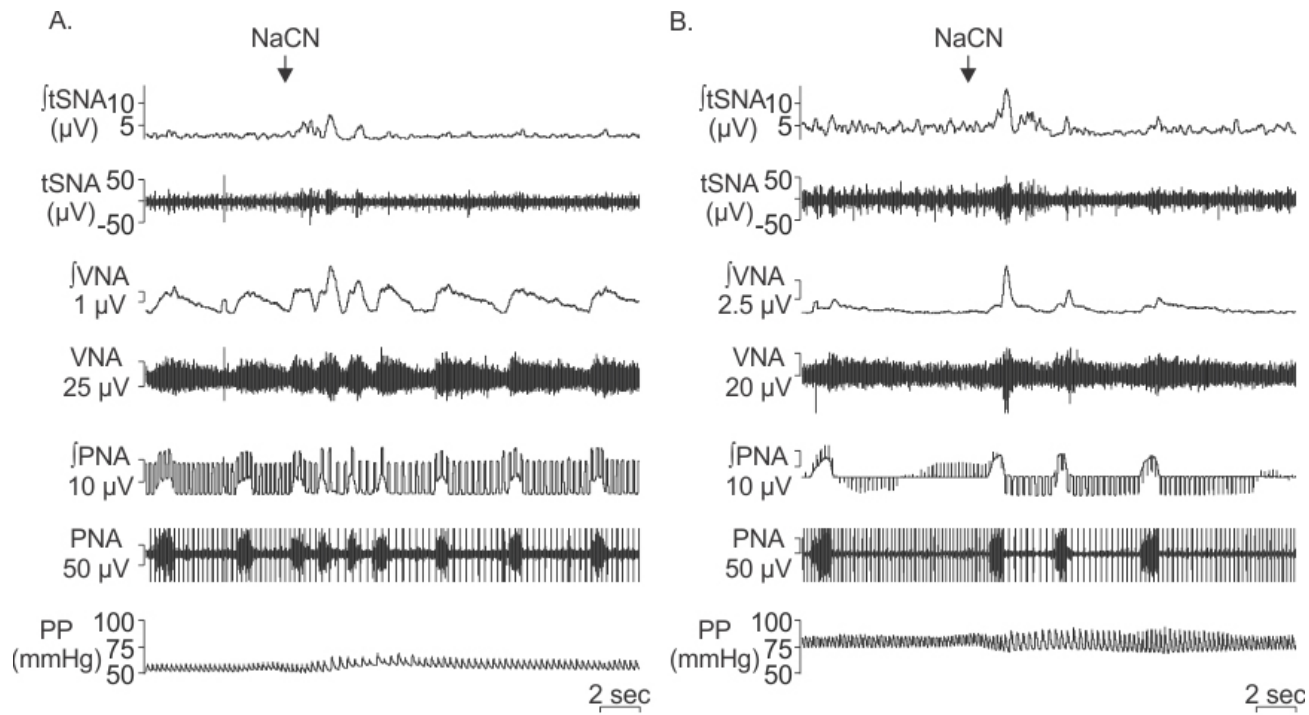
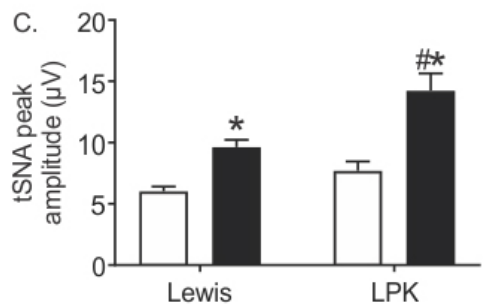
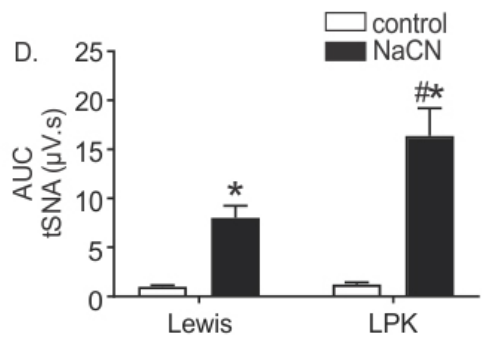

Figure 6

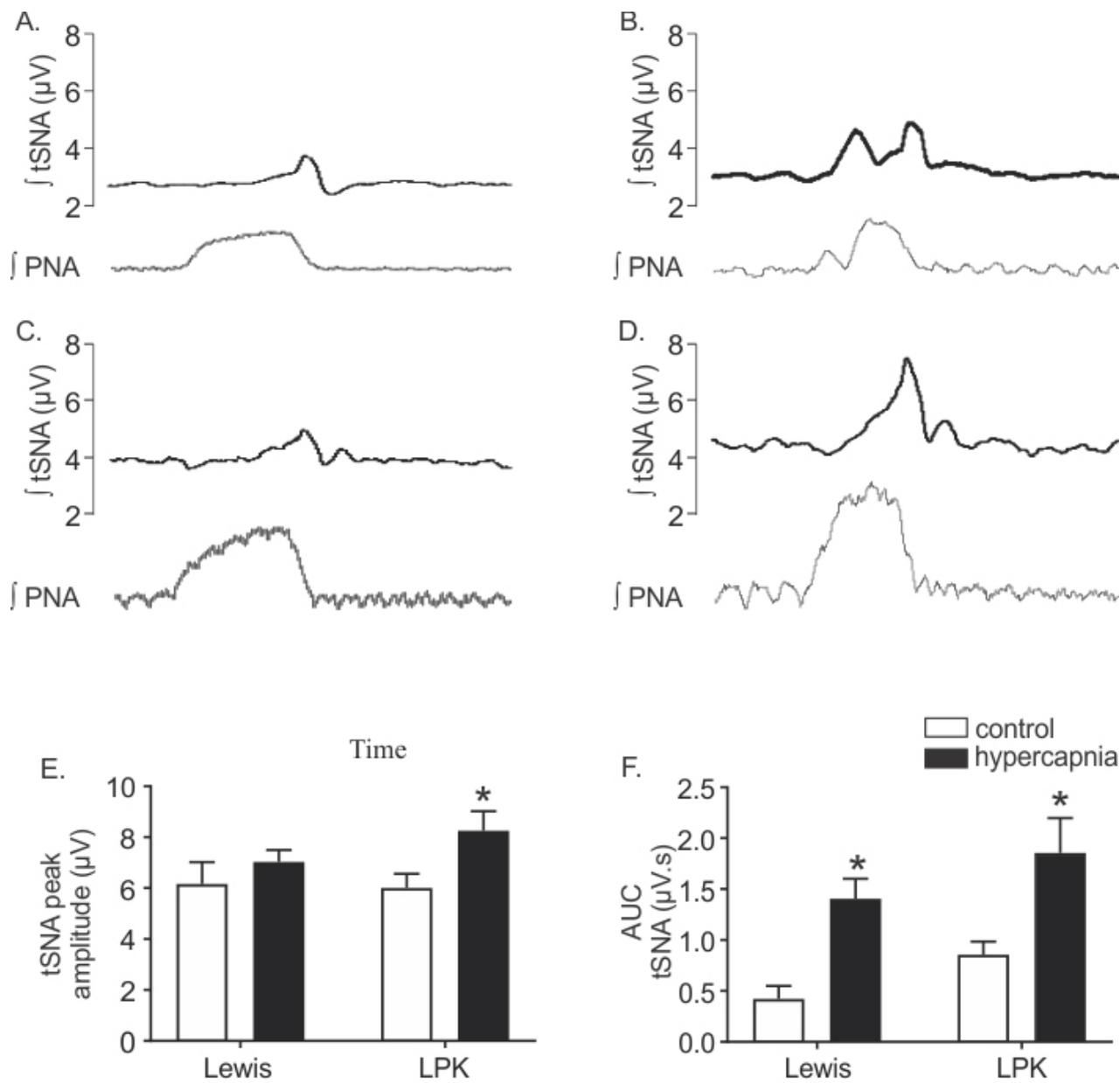

Figure 7

*Table 1: Baseline cardiorespiratory function in adult Lewis and LPK rats*

|  | Lewis | LPK |
| --- | --- | --- |
| MAP (mmHg) | 90 ± 4 | 125 ± 10* |
| SBP (mmHg) | 117 ± 8 | 194 ± 22* |
| DBP (mmHg) | 75 ± 3 | 99 ± 10* |
| PP (mmHg) | 41 ± 6 | 87 ± 18* |
| HR (bpm) | 453 ± 9 | 480 ± 3* |
| PNA amplitude (μV) | 18.3 ± 2.1 | 24.9 ± 5.3 |
| PNA frequency (cycles/min) | 36 ± 1 | 45 ± 1* |
| PNA duration (s) | 0.82 ± 0.1 | 0.72 ± 0.01 |
| MPA | 674.9 ± 86.03 | 1125 ± 232 |
| sSNA (μV) | 3.1 ± 0.5 | 7.9 ± 1.3* |
| rSNA (μV) | 5.2 ± 0.4 | 8.9 ± 0.6* |

Measures of cardiorespiratory function in Lewis and LPK rats. MAP: mean arterial pressure, SBP: systolic blood pressure, DBP: diastolic blood pressure, PP: pulse pressure, HR: heart rate, bpm: beats per min, PNA: phrenic nerve amplitude, MPA: minute phrenic activity, rSNA: renal sympathetic nerve activity, sSNA: splanchnic sympathetic nerve activity. Results are expressed as mean ± SEM. *P<0.05 between the Lewis and LPK as determined by Student's t-test. *n* = 9 Lewis and n = 8 LPK excepting rSNA where n = 5 LPK and 6 Lewis.

*Table 2: RespSNA parameters in splanchnic and renal sympathetic nerves under control conditions.*

|  |  | Lewis | LPK |
|---|---|---|---|
| PA (μV) | splanchnic | 4.1 ± 0.8 | 8.8 ± 1.1* |
|  | renal | 8.7 ± 0.6 | 11.02 ± 0.8* |
| Duration (sec) | splanchnic | 0.8 ± 0.07 | 0.99 ± 0.02 |
|  | renal | 0.8 ± 0.05 | 0.96 ± 0.05 |
| AUC (μV.s) | splanchnic | 3.5 ± 0.5 | 8.7 ± 1.1* |
|  | renal | 7.1 ± 0.2 | 10.6 ± 1.1* |

Measures of respSNA (phrenic triggered) parameters for integrated sSNA and rSNA in Lewis and LPK rats. PA: peak amplitude, AUC: area under curve. Results are expressed as mean ± SEM. *$P<0.05$ between the Lewis and LPK as determined by Student's t-test. sSNA data: *n* = 9 Lewis and n = 8 LPK, excepting rSNA where n = 5 LPK and 6 Lewis.

*Table 3: Effects of peripheral and central chemoreceptor stimulation on cardiorespiratory parameters in adult Lewis and Lewis Polycystic Kidney (LPK) rats*

|  |  | Lewis (n = 9) | LPK (n = 8) |
|---|---|---|---|
| Hypoxia | Δ MAP (mmHg) | 4 ± 4 | 21 ± 5 * |
|  | Δ SBP (mmHg) | 5 ± 5 | 34 ± 11 * |
|  | Δ DBP (mmHg) | 4 ± 4 | 16 ± 4 |
|  | Δ PP (mmHg) | 1 ± 2 | 18 ± 7 * |
|  | Δ HR (bpm) | 12 ± 2 | 10 ± 1 |
|  | Δ PNA amplitude (µV) | 10.9 ± 3.6 | 9.8 ± 3.9 |
|  | Δ PNA duration (sec) | 0.13 ± 0.02 | 0.07 ± 0.01 |
|  | Δ PNA frequency (cycles/min) | 5 ± 2 | -2 ± 2* |
|  | Δ MPA | 536.6 ± 152.5 | 300.7 ± 66.9 |
| Hypercapnea | Δ MAP (mmHg) | 19 ± 3 | 16 ± 4 |
|  | Δ SBP (mmHg) | 22 ± 4 | 29 ± 8 |
|  | Δ DBP (mmHg) | 18 ± 3 | 13 ± 3 |
|  | Δ PP (mmHg) | 4 ± 2 | 16 ± 6 |
|  | Δ HR (bpm) | -8 ± 1 | -4 ± 1 * |

| | | |
|---|---|---|
| Δ PNA amplitude (μV) | 8.7 ± 2.7 | 9.1 ± 1.6 |
| Δ PNA duration (sec) | 0.15 ± 0.02 | 0.11 ± 0.02 |
| Δ PNA frequency (cycles/min) | -3 ± 1 | -4 ± 1 |
| Δ MPA | 178.5 ± 74.08 | 293.9 ± 71.04 |

Delta change in phrenic nerve activity (PNA) and blood pressure (mmHg) in hypoxia (ventilated with only room air) or hypercapnia (ventilated with 5% $CO_2$ with 95% $O_2$) when switched from control condition (ventilated with oxygen enriched room air) in adult Lewis and Lewis Polycystic Kidney (LPK) rats under urethane anaesthesia. MAP: mean arterial pressure, SBP: systolic blood pressure, DBP: diastolic blood pressure, PP: pulse pressure; HR: heart rate, MPA: minute phrenic activity. Results are expressed as mean ± SEM. *P<0.05 between the Lewis and LPK as determined by Student's t-test. Δ = Delta change in, n = number of animals per group